\RequirePackage{fix-cm}
\documentclass[smallextended]{svjour3}       
\smartqed  
\usepackage{graphicx}
\usepackage{mathptmx}      
\usepackage{latexsym}
\usepackage{array, float, booktabs, amsmath, amssymb, geometry, mathtools, subcaption, relsize, bbm}
\allowdisplaybreaks
\usepackage[unicode]{hyperref}
\hypersetup{
    colorlinks=true,
    linkcolor=red,
    filecolor=magenta,      
    urlcolor=cyan,
}
\usepackage[font=small,labelfont=bf]{caption}

\begin{document}

\title{A new perturbative solution to the motion around triangular Lagrangian points in the elliptic restricted three-body problem
}
%


\titlerunning{Analytic solution to the ERTBP}        

\author{B\'alint Boldizs\'ar         \and
        Tam\'as Kov\'acs \and
        J\'ozsef Vany\'o 
}


\institute{B. Boldizs\'ar \at
              Institute of Physics, E\"otv\"os Iniversity, Budapest, Hungary \\
              \email{bolbalaa@caesar.elte.hu}           
            \and
            T. Kov\'acs \at
              Institute of Physics, E\"otv\"os Iniversity, Budapest, Hungary \\
              \email{tkovacs@general.elte.hu}           
            \and
            J. Vany\'o \at
              Department of Physics, Eszterházy Károly College, Eger, Hungary\\
              \email{vanyo.jozsef@uni-eszterhazy.hu}
}

\date{Received: date / Accepted: date}

\maketitle
\begin{abstract}
The equations of motion of planar elliptic restricted three body problem are transformed to four decoupled Hill's equations. By using the Floquet theorem analytic solution to the oscillator equations with time dependent periodic coefficients are presented. We show that the new analytic approach is valid for system parameters $0 < e \leq 0.05$ and $0 < \mu \leq 0.01$ where $e$ denotes the eccentricity of primaries while $\mu$ is the mass parameter, respectively. We also clarify the transformation details that provide the applicability of the method.

\end{abstract}

\section{Introduction}
\label{intro}
In the era of exoplanets and specifically designed space missions, the co-orbital motion in the vicinity of the equilateral points $L_4$ and $L_5$ became again the focus of attention. Since the seminal work of Szebehely \cite{Szebehely1967} the orbits near the libration points have been discussed extensively by the community. Analytic description of the Trojan-like resonant dynamics in the elliptic case of restricted three-body problem (ERTBP) is based mainly on averaged motion. Érdi \cite{Erdi1977,Erdi1978} showed the perturbation effects up to second order in Jupiter's eccentricity, perihelion and ascending node precession by using a three-parameter expansion. Morais \cite{Morais2001} considered an averaged disturbing potential to describe the secular variation of the Trojans' orbital elements in case of an oblate primary. Recently, \cite{Robutel2016} and \cite{Paez2016} investigated the co-orbital resonance based on Hamiltonian formalism whereby the fast angles had been averaged out. These latter analytical studies are also capable to locate higher-order resonances as well as very slow secular frequencies.

It has been demonstrated \cite{Tschauner1971,Erdi1974,Meire1981,Matas1982} that the coupled equations of the ERTBP can be written in the form of independent ordinary differential equations with variable coefficients.  The primary goal of these studies is to explore the stability map of eccentricity--mass parameter dated back to \cite{Danby1964}. Interestingly, the analysis given by \cite{Erdi1977} [Eq.~(24)] and \cite{Robutel2016} [Eq.~(56)] also terminates at a pendulum-like equation, however, they do not attempt to solve it by classical techniques such as Floquet theorem \cite{Lichtenberg1983}.
Here we propose a detailed derivation of Hill's equations of ERTBP and make a comprehensive analysis of their applicability which is still out of literature. Furthermore, analytic expressions for the solution of Hill's equations are given in the regime of moderate eccentricities and mass parameter with good agreement of numerical calculations.
  
\section{Basic context}
\label{sec:context}

In this paper we mainly follow the notations used in e.g. \cite{Tschauner1971,Meire1980,Meire1982}. Motion around the $L_4$ and $L_5$ Lagrangian points is determined by the coupled differential equations \cite{Szebehely1967}
\begin{align}\label{e:de1}
&x''-2y' = rc_1x,\\\label{e:de2}
&y''+2x' = rc_2y,
\end{align}
where the notations are
\begin{align}
&r = \frac{1}{1+e\cos(v)},\quad \mu = \frac{m_2}{m_1+m_2},\quad g=3\mu(1-\mu)\quad\textnormal{and} \quad c_i =\frac 32 (1+(-1)^i\sqrt{1-g})\quad (i=1,2).
\end{align}
Here $'$ denotes the derivation with respect to the true anomaly $v$.

\subsection{Hill's equations}
We will show, that Eqs.~\eqref{e:de1}-\eqref{e:de2} with a suitable transformation can be rewritten to four decoupled second order differential equations. Let us introduce $y_1,y_2,y_1^*,y_2^*$ with the following transformation
\begin{align}\label{e:p_traf}
\begin{pmatrix}
x \\ y \\ x' \\ y'
\end{pmatrix}
=
\begin{pmatrix}
\mathbbm 1_2 & \mathbbm 1_2 \\
\mathbf P_1 & \mathbf P_2
\end{pmatrix}
\begin{pmatrix}
y_1 \\ y_2 \\ y_1^* \\ y_2^*,
\end{pmatrix},
\end{align}
where $\mathbbm 1_2$ is the 2-dimensional identity matrix, and furthermore $\mathbf P_1$ and $\mathbf P_2$ are introduced as
\begin{align}\label{e:p_traf_der}
\begin{pmatrix}
y_1' \\ y_2' \\ {y_1^*}' \\ {y_2^*}'
\end{pmatrix}
=
\begin{pmatrix}
\mathbf P_1 & \mathbf 0 \\
\mathbf 0 & \mathbf P_2
\end{pmatrix}
\begin{pmatrix}
y_1 \\ y_2 \\ y_1^* \\ y_2^*
\end{pmatrix}
\end{align}
relation stands. Let us use the temporary notations $\tilde x = (x,y)$, $\tilde y_1 = (y_1, y_2)$ and $\tilde y_2 = (y_1^*, y_2^*)$. The elements of $\mathbf P_i$ ($i = 1,2$) matrices can be gained by using the following identities
\begin{align}
\begin{pmatrix}
\tilde x \\ \tilde  x'
\end{pmatrix}
=
\begin{pmatrix}
\mathbbm 1_2 & \mathbbm 1_2 \\
\mathbf P_1 & \mathbf P_2
\end{pmatrix}
\begin{pmatrix}
\tilde y_1 \\ \tilde y_2
\end{pmatrix}
=
\begin{pmatrix}
\tilde y_1 + \tilde y_2 \\
\mathbf P_1 \tilde y_1 +\mathbf P_2 \tilde y_2
\end{pmatrix},
\end{align}
from which it simply follows, that
\begin{align}
\begin{pmatrix}
\tilde x' \\ \tilde x''
\end{pmatrix}
=
\begin{pmatrix}
\tilde y_1'+\tilde y_2'\\
\mathbf P_1' \tilde y_1+\mathbf P_1 \tilde y_1' + \mathbf P_2' \tilde y_2 + \mathbf P_2 \tilde y_2'
\end{pmatrix}
=
\begin{pmatrix}
\mathbf 0 & \mathbbm 1_2 \\
r\mathbf C & 2\mathbf D
\end{pmatrix}
\begin{pmatrix}
\tilde x \\ \tilde x'
\end{pmatrix}
=
\begin{pmatrix}
\mathbf 0 & \mathbbm 1_2 \\
r\mathbf C & 2\mathbf D
\end{pmatrix}
\begin{pmatrix}
\mathbbm 1_2 & \mathbbm 1_2 \\
\mathbf P_1 & \mathbf P_2
\end{pmatrix}
\begin{pmatrix}
\tilde y_1 \\ \tilde y_2
\end{pmatrix},
\end{align}
where $\mathbf C = \begin{pmatrix}
c_1 & 0 \\ 0 & c_2
\end{pmatrix}$ and $\mathbf D = \begin{pmatrix}
0 & 1 \\ -1 & 0.
\end{pmatrix}$. It can be recognized, that
\begin{align}\label{e:mtx_de}
\begin{pmatrix}
\tilde y_1'+\tilde y_2'\\
\mathbf P_1' \tilde y_1+\mathbf P_1 \tilde y_1' + \mathbf P_2' \tilde y_2 + \mathbf P_2 \tilde y_2'
\end{pmatrix}
=
\begin{pmatrix}
\mathbf P_1 & \mathbf P_2 \\
\mathbf P_1'+\mathbf P_1^2 & \mathbf P_2' + \mathbf P_2^2
\end{pmatrix}
\begin{pmatrix}
\tilde y_1 \\ \tilde y_2
\end{pmatrix}
\quad\Rightarrow\quad \mathbf P_i'+\mathbf P_i^2 = r\mathbf C + 2\mathbf D \mathbf P_i,
\end{align}
which is a Riccatti-type matrix differential equation. Based on Tschauner's argument~\cite{Tschauner1971} the following matrix elements satisfy Eq.~\eqref{e:mtx_de}
\begin{align}
p_{11}^{(i)} &= -\frac 12 re\sin(v)(1+ke\cos(v)),\\
p_{12}^{(i)} &= r\left(a_2^{(i)}+e\cos(v)-\frac 14 ke^2\cos(2v) \right),\\
p_{21}^{(i)} &= -r\left(a_1^{(i)}+e\cos(v)+\frac 14 ke^2\cos(2v) \right),\\
p_{22}^{(i)} &= -\frac 12 re\sin(v)(1-ke\cos(v)),
\end{align}
where
\begin{align}\hspace{-3mm}
k = \frac{1}{\sqrt{1-g}},\quad c = \sqrt{1{-}9g{+}2e^2{+}k^2e^4},\quad a_1^{(i)} = \frac 14(2c_1 {+} 1 {+}({-}1)^i c),\quad a_2^{(i)} = \frac 14(2c_2 {+} 1 {+}({-}1)^i c)
\end{align}
notations are introduced. As all elements of $\mathbf P_i$ have a multiplicative factor of $r$, therefore we define the following $q=p/r$ quantities as
\begin{align}
q_{11}^{(i)} &= -\frac 12 e\sin(v)(1+ke\cos(v)) & \Rightarrow\quad {q_{11}^{(i)}}' &= -\frac 12(e\cos(v)+ke^2\cos(2v)),\\ \label{e:q11}
q_{12}^{(i)} &= \left(a_2^{(i)}+e\cos(v)-\frac 14 ke^2\cos(2v) \right) & \Rightarrow\quad {q_{12}^{(i)}}' &= 2q_{22}^{(i)},\\ \label{e:q12}
q_{21}^{(i)} &= -\left(a_1^{(i)}+e\cos(v)+\frac 14 ke^2\cos(2v) \right) & \Rightarrow\quad {q_{21}^{(i)}}' &= -2q_{11}^{(i)},\\ \label{e:q21}
q_{22}^{(i)} &= -\frac 12 e\sin(v)(1-ke\cos(v)) & \Rightarrow\quad {q_{22}^{(i)}}' &= -\frac 12(e\cos(v)-ke^2\cos(2v)). \\ \label{e:q22}
\end{align}
Let $\det \mathbf Q^{(i)}$ be the determinant of the matrix with the above elements. It can be shown
\begin{align}
\det \mathbf Q^{(i)} = \frac{1}{2r} \left((-1)^ic+1+3e\cos(v) \right).
\end{align}
According to Eq.~\eqref{e:p_traf_der} $\tilde y_i' = \mathbf P_i \tilde y_i$, from which the $\tilde y_i'' = \mathbf P_i'\tilde y_i + \mathbf P_i \tilde y_i' = (2\mathbf D\mathbf P + r\mathbf C)\tilde y_i$ relation follows. Consequently,
\begin{align}\label{e:y1dd}
y_1'' &= (2p_{21}^{(1)}+rc_1)y_1 + 2p_{22}^{(1)} y_2 = \frac{1}{q_{12}^{(1)}}(q_{12}^{(1)}rc_1-2r\det \mathbf Q^{(1)})y_1 + \frac{2q_{22}^{(1)}}{q_{12}^{(1)}}y_1',\\\label{e:y2dd}
y_2'' &= (-2p_{12}^{(1)}+rc_2)y_2 - 2p_{11}^{(1)} y_1 = \frac{1}{q_{21}^{(1)}}(q_{21}^{(1)}rc_2+2r\det \mathbf Q^{(1)})y_2 - \frac{2q_{11}^{(1)}}{q_{21}^{(1)}}y_2',
\end{align}
where we used the relations
\begin{align}\label{e:y2y1d}
y_1 = \frac{y_2'-p_{22}^{(1)}y_2}{p_{21}^{(1)}}\quad\textnormal{and}\quad y_2 = \frac{y_1'-p_{11}^{(1)}y_1}{p_{12}^{(1)}}.
\end{align}
Similar arguments are true for $y_1^*$ and $y_2^*$ with elements of matrix $\mathbf P_2$.

Considering the general form of the ordinary differential equation $y'' + a(v) y' + b(v) y = 0,$ the transformation $y=\xi(v) \exp(-1/2 \int_0^v a(x) \mathrm dx)$ eliminates the first order derivative term $y'.$ Applying this conversion to Eqs.~(\ref{e:y1dd}) and \eqref{e:y2dd} we can introduce the following transformations
\begin{align}\label{e:y1}
y_1 &= \sqrt{|q_{12}^{(1)}|}\xi_1, &y_1' &= \frac{q_{22}^{(1)}}{\sqrt{|q_{12}^{(1)}}|}\xi_1+\sqrt{|q_{12}^{(1)}|}\xi_1', &y_1'' &= \frac{{q_{22}^{(1)}}'|q_{12}^{(1)}|-{q_{22}^{(1)}}^2}{{|q_{12}^{(1)}|}^{3/2}}\xi_1+\frac{2q_{22}^{(1)}}{\sqrt{|q_{12}^{(1)}}|}\xi_1'+\sqrt{|q_{12}^{(1)}|}\xi_1'', \\ \label{e:y1*}
y_1^* &= \sqrt{|q_{12}^{(2)}|}\xi_2, &{y_1^*}' &= \frac{q_{22}^{(2)}}{\sqrt{|q_{12}^{(2)}}|}\xi_2+\sqrt{|q_{12}^{(2)}|}\xi_2', &{y_1^*}'' &= \frac{{q_{22}^{(2)}}'|q_{12}^{(2)}|-{q_{22}^{(2)}}^2}{{|q_{12}^{(2)}|}^{3/2}}\xi_2+\frac{2q_{22}^{(2)}}{\sqrt{|q_{12}^{(2)}}|}\xi_2'+\sqrt{|q_{12}^{(2)}|}\xi_2''.
\end{align}
By using equations above, \eqref{e:y1dd} and \eqref{e:y2dd} become the differential equations of harmonic oscillators with periodic coefficient. These equations are also known as Hill's equation
\begin{align}\label{e:hill1}
\hspace{-3mm}\xi_1'' &+ J_1(v)\xi_1 = 0,\quad\textnormal{where}\quad J_1(v) = -\left(rc_1+2-\frac{3r\det\mathbf Q^{(1)}+c_2}{q_{12}^{(1)}}+\frac{3{q_{22}^{(1)}}^2}{{q_{12}^{(1)}}^2}\right),\quad y_1 = \sqrt{|q_{12}^{(1)}|}\xi_1,\\
\label{e:hill3}
    \xi_3'' &+ J_3(v)\xi_3 = 0,\quad\textnormal{where}\quad J_3(v) = -\left(rc_2+2+\frac{3r\det\mathbf Q^{(1)}+c_1}{q_{21}^{(1)}}+\frac{3{q_{11}^{(1)}}^2}{{q_{21}^{(1)}}^2}\right),\quad y_2 = \sqrt{|q_{21}^{(1)}|}\xi_3
\end{align}

The corresponding transformations (not presented) can also be carried out for $y_1^*$ and $y_2^*$ yielding the following Hill's equations 
\begin{align}
\label{e:hill2}
\hspace{-3mm}\xi_2'' &+ J_2(v)\xi_2 = 0,\quad\textnormal{where}\quad J_2(v) = -\left(rc_1+2-\frac{3r\det\mathbf Q^{(2)}+c_2}{q_{12}^{(2)}}+\frac{3{q_{22}^{(2)}}^2}{{q_{12}^{(2)}}^2}\right),\quad y_1^* = \sqrt{|q_{12}^{(2)}|}\xi_2,\\
    \label{e:hill4}
    \xi_4'' &+ J_4(v)\xi_4 = 0,\quad\textnormal{where}\quad J_4(v) = -\left(rc_2+2+\frac{3r\det\mathbf Q^{(2)}+c_1}{q_{21}^{(2)}}+\frac{3{q_{11}^{(2)}}^2}{{q_{21}^{(2)}}^2}\right),\quad y_2^* = \sqrt{|q_{21}^{(2)}|}\xi_4.
\end{align}
In Eqs.~\eqref{e:hill1}-\eqref{e:hill4} $J_i$ ($i=1\dots 4$) are periodic coefficients with period of $2\pi$. Square root of the coefficients gives the frequency of the oscillator. We can obtain the original Cartesian coordinates by using Eq.~\eqref{e:p_traf}, thus, $x,y$ coordinates can be calculated as $x=y_1+y_1^*$ and $y = y_2+y_2^*$. 

\begin{figure}[htb]
\centering
\includegraphics[width=0.5\textwidth]{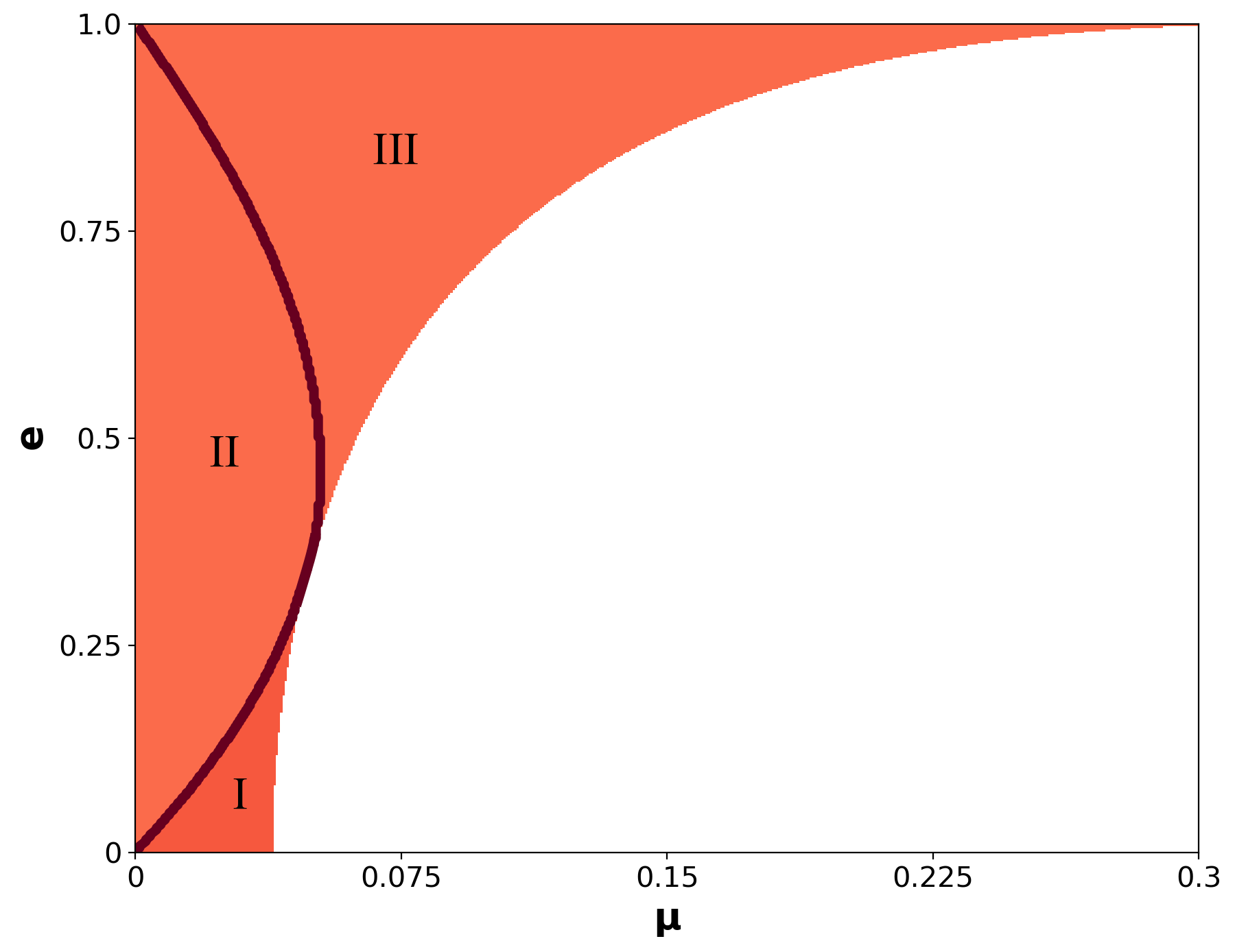}
\caption{Shaded region of $\mu-e$ parameter plane describes where Hill's equations might have real solutions. Along the black solid lines between domains I, II, and III. condition $q_{21}^{(i)}=0$ holds. That is Eqs.~\eqref{e:hill2} and \eqref{e:hill4} have no solutions. No stable solution exists in white part.}
\label{f:q_12_condition}
\end{figure}

It is clear from the coefficients $J_i$ that Eqs.~\eqref{e:hill1}-\eqref{e:hill4} do not have solutions if $q_{jk}^{(l)}(\mu,e)=0$~($j=1 \text{ or } 2$, $k=1 \text{ or } 2$ and $l= 1 \text{ or } 2,$ see Eqs.~\eqref{e:q11}-\eqref{e:q22}). The forbidden parameter pairs ($\mu,e$) as solid lines are depicted in Fig.~\ref{f:q_12_condition}. We note that $q_{12}^{(1)}$ and $q_{12}^{(2)}$ associated to Eqs.~\eqref{e:hill1} and \eqref{e:hill2} do not take zero value anywhere in the shaded region. However, the black solid line between domain I and II corresponds to those ($\mu,e$) pairs where $q_{21}^{(1)}=0.$ Similarly, $q_{21}^{(2)}=0$ along the line between the regions II and III.

\begin{figure}[htb]
\centering
\includegraphics[width=0.7\textwidth]{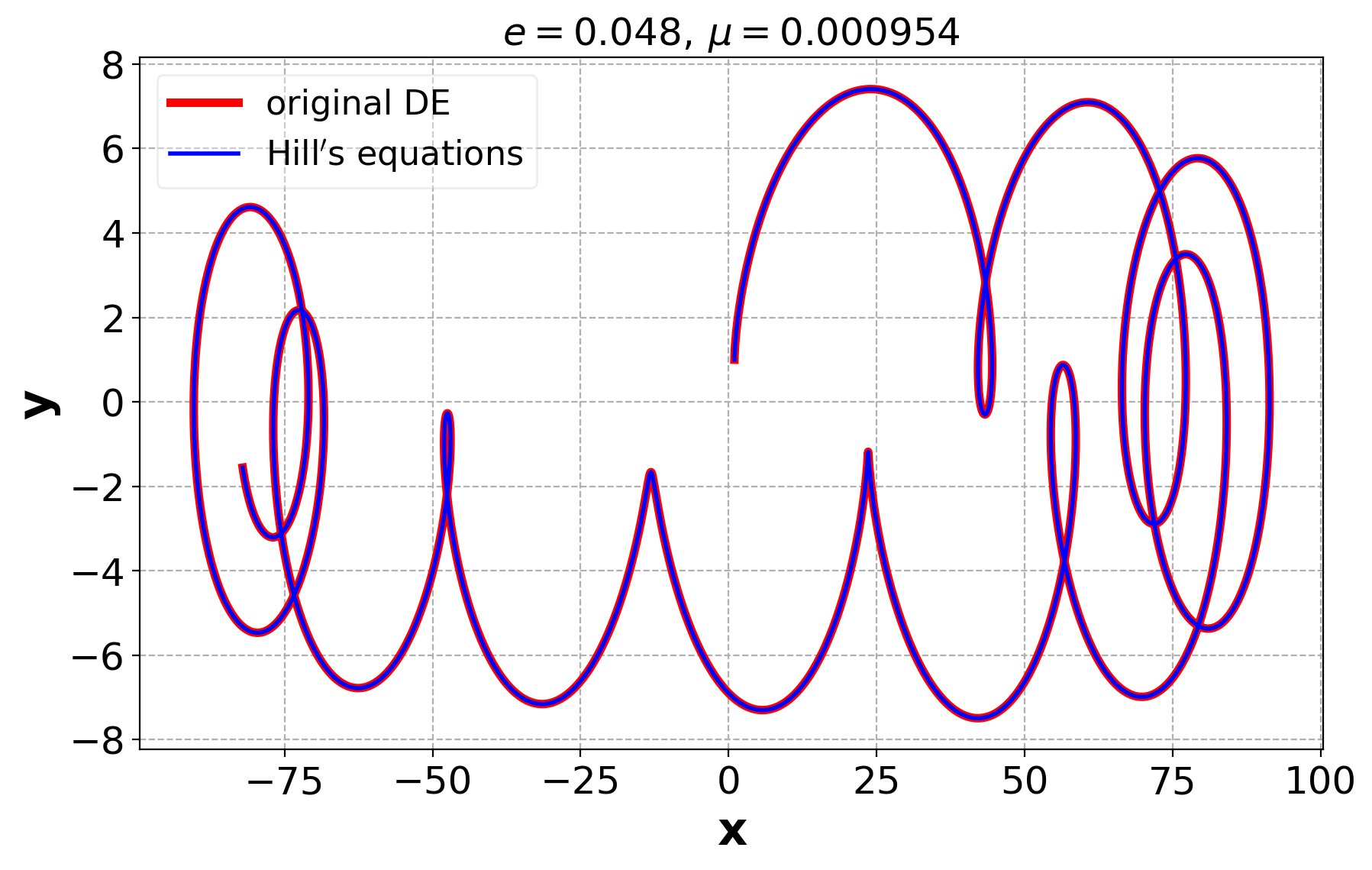}
\caption{Numerical solutions around the $L_4$ and $L_5$ points. The initial conditions and the parameters are $x_0 = 1$, $y_0 = 1$, $v_{x0} = 0$, $v_{y0} = 0$, $e=0.048$, $\mu=0.000954$ (the case of Jupiter), respectively.}
\label{f:DE_and_Hill_v}
\end{figure}

Transformations Eqs.~\eqref{e:y1}-\eqref{e:y1*} can be substituted into Eq.~\eqref{e:y2y1d}, from which we get
\begin{align}\label{e:y2y2*}
    y_2 = \left(\frac{q_{22}^{(1)}}{r{|q_{12}^{(1)}|}^{3/2}}-\frac{q_{11}^{(1)}}{\sqrt{|q_{12}^{(1)}|}}\right)\xi_1+\frac{1}{r\sqrt{|q_{12}^{(1)}|}}\xi_1'\quad\textnormal{and}\quad
    y_2^*= \left(\frac{q_{22}^{(2)}}{r{|q_{12}^{(2)}|}^{3/2}}-\frac{q_{11}^{(2)}}{\sqrt{|q_{12}^{(2)}|}}\right)\xi_2+\frac{1}{r\sqrt{|q_{12}^{(2)}|}}\xi_2'.
\end{align}
Doing this Eq.~\eqref{e:y2y2*} reduces the four Hill’s equations to two. Thus Eqs.~\eqref{e:hill1} and \eqref{e:hill2} fully describe the problem\footnote{\,We note that any two equations can be selected but for practical reasons the pair of $\xi_1$ and $\xi_2$ is the best choice.}. In other words, Eqs.~\eqref{e:y2y2*} allows one to use safely the transformations \eqref{e:y1} and \eqref{e:y1*} to solve Hill's equations. Fig.~\ref{f:DE_and_Hill_v} depicts the trajectory for $e=0.048$, $\mu=0.000954$ (the case of Jupiter). The solution of Eqs.~\eqref{e:de1}-\eqref{e:de2} and Eqs.~\eqref{e:hill1}-\eqref{e:hill2} originating from the appropriate initial conditions perfectly overlap. This means, that the transformations Eqs.~\eqref{e:y1}-\eqref{e:y1*} lead to the same result, therefore, Hill's equations can be applied to solve the equations of motion around the $L_4$ and $L_5$ points.

\section{Perturbative solution}
In this section we give the perturbative solution of the differential equations.
Hill's equations (Eqs.~\eqref{e:hill1}-\eqref{e:hill2}), as they are second order differential equations with periodic coefficients, can be solved by Floquet theorem~\cite{Hagel1992}. We seek the solution in the form of
\begin{align}
\xi(v) = a w(v) \cos(\psi(v)+b),
\end{align} 
where $w(v)$ is the so-called Floquet function, which has the same period as $\xi(v)$. Constants $a$ and $b$ are determined by the initial conditions. Since the derivation for both $\xi_1$ and $\xi_2$ are the same, we omit the indices in the rest part of the paper. Let us rewrite Eqs.~\eqref{e:hill1}-\eqref{e:hill2} to $w(v)$ and $\psi(v)$
\begin{align}
\xi'(v) &= a w'(v)\cos(\psi(v)+b)-aw(v)\sin(\psi(v)+b)\psi'(v),\\
\begin{split}
\xi''(v) &= a w''(v)\cos(\psi(v)+b)-2aw'(v)\sin(\psi(v)+b)\psi'(v)-aw(v)\cos(\psi(v)+b){\psi'}^2(v)\\&-aw(v)\sin(\psi(v)+b)\psi''(v).
\end{split}
\end{align} 
The differential equations above split into two parts with the coefficients of $\sin$ and $\cos$
\begin{align}\label{e:wprime}
&w''(v)-w(v){\psi'}^2(v)+J(v)w(v) = 0,\\\label{e:psiprime}
&2w'(v)\psi'(v)+w(v)\psi''(v) = 0.
\end{align}
From Eq.~\eqref{e:psiprime} we obtain
\begin{align}
2\frac{w'(v)}{w(v)} = -\frac{\psi''(v)}{\psi'(v)} \quad\Rightarrow\quad \frac{\mathrm d}{\mathrm d v}\log(w^2(v)) = \frac{\mathrm d}{\mathrm d v} \log \left(\frac{1}{\psi'(v)}\right)\quad\Rightarrow\quad \psi'(v) = \frac{C}{w^2(v)},
\end{align}
where $C$ is a constant. With this equation Eq.~\eqref{e:wprime} becomes
\begin{align}\label{e:wprime1}
w''+J(v)w(v)-\frac{C^2}{w^3(v)}=0.
\end{align}
Now we are looking for the solution of $w(v)$ in a third order Taylor series in the eccentricity $e$
\begin{align}
w(v) = w^{(0)}(v)+ew^{(1)}(v)+e^2w^{(2)}(v)+e^3 w^{(3)}(v)+\mathcal O (e^4).
\end{align}

\subsection{Taylor series of $J_1(v)$ and $J_2(v)$}\label{sec:series}
The periodic coefficients to be solved have complicated forms, therefore, the solution can be obtained by a third order Taylor expansion in the eccentricity. Let us first utilize $J_1$
\begin{align}
\hspace{-3mm}\resizebox{.95\hsize}{!}{$\displaystyle J_1(v) {=} {-}\Bigg\{c_1 r {+}2 {-}\frac{\frac 32 (1{-}c+3e\cos(v)){+}c_2}{\frac 14 (2c_2{+}1{-}c){+}e\cos(v){-}\frac 14 ke^2\cos(2v)}{+}3\left(\frac{{-}\frac 12 e\sin(v)(1{-}ke\cos(v))}{\frac 14(2c_2{+}1{-}c){+}e\cos(v){-}\frac 14 ke^2\cos(2v)}\right)^2\Bigg\},$}
\end{align}
by using the earlier introduced notations. Useful expressions will be $\lambda\equiv \sqrt{1-9g}$ and $B~{\equiv}~(2c_2~{+}~1~{-}~\lambda)^{-1}$. Third order Taylor expansion of $J_1$ is then
\begin{align*}
J_1(v;e,\mu) &= \alpha_1+\beta_1\cos(v)e+(\gamma_1+\delta_1\cos(2v))e^2+(\varepsilon_1\cos(v)+\eta_1\cos(3v))e^3+\mathcal O(e^4),
\end{align*}
where
\begin{align*}
\alpha_1 &= -c_1-2+B(6-6\lambda+4c_2),\\
\beta_1 &= c_1+18B+8B^2(-3+3\lambda-2c_2),\\
\gamma_1 &= \frac{B^2}{\lambda}(6-12\lambda+4c_2)-\frac{6B}{\lambda}-\frac{c_1}{2}-4B^3(10c_2-3+3\lambda),\\
\delta_1 &= B^2 k \left(6-6\lambda+4c_2+\frac 6k\right)-\frac{c_1}{2}-4B^3(10c_2-3+3\lambda),\\
\varepsilon_1 &= \frac{B^4}{\lambda}\Bigg\{\frac{20c_2-6+30\lambda+10c_2\lambda k-3\lambda k+3\lambda^2 k}{B}+\frac{6k\lambda}{B^2}+64c_2^2+32c_2+\tag{\stepcounter{equation}\theequation}\\
&+208c_2\lambda-32c_2k\lambda-12k\lambda+24k-216gk+32c_2k-288kc_2g-12k\lambda^3-16c_2^2k\lambda-\\&-72\lambda+72-648g\Bigg\}+\frac{3c_1}{4},\\
\eta_1 &= \frac{B^4}{\lambda}\Bigg\{\frac{10c_2\lambda k-3\lambda k+3 \lambda^2 k+24\lambda}{B}+\frac{6k\lambda}{B^2}-32c_2k\lambda-12k\lambda+24k-216gk+\\&+32c_2k-288c_2 gk-12k\lambda^3-16c_2^2k\lambda+80\lambda c_2-24\lambda+24-216g \Bigg\}+\frac{c_1}{4}.
\end{align*}
The expression of the other periodic coefficient $J_2$ is
\begin{align}
\hspace{-3mm}\resizebox{.95\hsize}{!}{$J_2(v) {=} {-}\Bigg\{c_1 r {+}2 {-}\frac{\frac 32 (1{+}c{+}3e\cos(v)){+}c_2}{\frac 14 (2c_2{+}1{+}c){+}e\cos(v){-}\frac 14 ke^2\cos(2v)}{+}3\left(\frac{{-}\frac 12 e\sin(v)(1{-}ke\cos(v))}{\frac 14(2c_2{+}1{+}c){+}e\cos(v){-}\frac 14 ke^2\cos(2v)}\right)^2\Bigg\}.$}
\end{align}
To calculate the Taylor series of $J_2$ (again up to third order in $e$) we use $D\equiv (2c_2+1+\lambda)^{-1}$. Then $J_2$ becomes
\begin{align*}
J_2(v;e,\mu) &= \alpha_2+\beta_2\cos(v)e+(\gamma_2+\delta_2\cos(2v))e^2+(\varepsilon_2\cos(v)+\eta_2\cos(3v))e^3+\mathcal O(e^4),
\end{align*}
where
\begin{align*}
\alpha_2 &= -c_1-2+D(6+6\lambda+4c_2),\\
\beta_2 &= c_1+18D-8D^2(3+3\lambda+2c_2),\\
\gamma_2 &= -\frac{D^2}{\lambda}(6+12\lambda+4c_2)+\frac{6D}{\lambda}-\frac{c_1}{2}+4D^3(-10c_2+3+3\lambda),\\
\delta_2 &= D^2 k \left(6+6\lambda+4c_2+\frac 6k\right)-\frac{c_1}{2}+4D^3(-10c_2+3+3\lambda)\tag{\stepcounter{equation}\theequation},\\
\varepsilon_2 &= \frac{D^4}{\lambda}\Bigg\{\frac{-20c_2+6+30\lambda+10c_2\lambda k-3\lambda k-3\lambda^2 k}{D}+\frac{6k\lambda}{D^2}-64c_2^2-32c_2+\\&+208c_2\lambda-32c_2k\lambda-12k\lambda-24k+216gk-32c_2k+288kc_2g-12k\lambda^3-16c_2^2k\lambda-\\&-72\lambda-72+648g\Bigg\}+\frac{3c_1}{4},\\
\eta_2 &= \frac{D^4}{\lambda}\Bigg\{\frac{10c_2\lambda k-3\lambda k-3 \lambda^2 k+24\lambda}{D}+\frac{6k\lambda}{D^2}-32c_2k\lambda-12k\lambda-24k+216gk-\\&-32c_2k+288c_2 gk-12k\lambda^3-16c_2^2k\lambda+80\lambda c_2-24\lambda-24+216g \Bigg\}+\frac{c_1}{4}.
\end{align*}
Let us write back the results of the Taylor expansions into Eq.~\eqref{e:wprime1}, and use the fact that
\begin{align}
\begin{split}
&\hspace{-3mm}\frac{1}{(w^{(0)}(v){+}ew^{(1)}(v){+}e^2w^{(2)}(v){+}e^3 w^{(3)}(v))^3}{=}\frac{1}{w^{(0)^3}(v)}{-}\frac{3w^{(1)}(v)}{w^{(0)^4}(v)}e{+}\frac{6w^{(1)^2}(v){-}3w^{(0)}(v)w^{(2)}(v)}{w^{(0)^5}(v)}e^2+\\&\hspace{-3mm}+\frac{-3w^{(0)^2}(v)w^{(3)}(v)+12w^{(0)}(v)w^{(1)}(v)w^{(2)}(v)-10w^{(1)^3}(v)}{w^{(0)^6}(v)}e^3+\mathcal O(e^4).
\end{split}
\end{align}
Then we can collect the terms for $e^0$, $e^1$, $e^2$ and $e^3$, thus 4 new differential equations can be obtained (also for $i=1,2$) for the terms of $w(v)$:
\begin{align}\label{e:diffeq0}
w^{(0)''}(v)&+w^{(0)}(v)\alpha -\frac{C^2}{w^{(0)^3}(v)} = 0,\\\label{e:diffeq1}
w^{(1)''}(v)&+w^{(0)}(v)\beta\cos(v)+w^{(1)}(v)\alpha +\frac{3C^2w^{(1)}(v)}{w^{(0)^4}(v)}=0,\\\label{e:diffeq2}
w^{(2)''}(v)&+w^{(0)}(v)\Big(\gamma+\delta\cos(2v) \Big)+w^{(1)}(v)\beta\cos(v)+w^{(2)}(v)\alpha -\frac{6C^2w^{(1)^2}(v)}{w^{(0)^5}} + \frac{3C^2w^{(2)}(v)}{w^{(0)^4}(v)}=0,\\\label{e:diffeq3}
\begin{split}
w^{(3)''}(v)&+w^{(0)}(v)\Big(\varepsilon\cos(v)+\eta\cos(3v) \Big)+w^{(1)}(v)\Big(\gamma+\delta\cos(2v)\Big) +w^{(2)}\beta\cos(v)+w^{(3)}(v)\alpha+\\&+\frac{3C^2w^{(3)}(v)}{w^{(0)^4}(v)}-\frac{12C^2w^{(1)}(v)w^{(2)}(v)}{w^{(0)^5}(v)}+\frac{10C^2w^{(1)^3}(v)}{w^{(0)^6}(v)}=0.
\end{split}
\end{align}
Again we note, that for all cases $w^{(j)}(v)=w^{(j)}(v+2\pi),\, (j=0,1,2,3)$, as also $\xi(v)=\xi(v+2\pi)$. It can be easily seen, that the unique solution for Eq.~\eqref{e:diffeq0} is:
\begin{align}
w^{(0)}(v)=\frac{C^{1/2}}{\alpha^{1/4}}\equiv w_{0,0}.
\end{align}
Differential equations \eqref{e:diffeq1}-\eqref{e:diffeq2}-\eqref{e:diffeq3} are second order linear differential equations, therefore the solution can be written up as the sum of the solution of the homogeneous equation ($w^{(j)}_h(v)$) and a particular solution of the inhomogeneous equation ($w^{(j)}_{ih}(v)$). Homogeneous part of Eq.~\eqref{e:diffeq1} is
\begin{align}
w^{(1)''}_h(v)+\left(\alpha+\frac{3C^2}{w_{0,0}^4}\right)w^{(1)}_h(v)=0,
\end{align}
which is a harmonic oscillator with frequency $\left(\alpha+\frac{3C^2}{w_{0,0}^4}\right)^{1/2}$, thus the solution of the equation is
\begin{align}
w^{(1)}_h(v) = K_1\sin\left(\sqrt{\alpha+\frac{3C^2}{w_{0,0}^4}}v\right)+K_2\cos\left(\sqrt{\alpha+\frac{3C^2}{w_{0,0}^4}}v\right),
\end{align}
where the constants $K_1$ and $K_2$ must be determined from the initial conditions. In order to fulfill the $2\pi$ periodicity of $w(v)$, the constants must be $K_1=K_2\equiv 0$. For the inhomogeneous solution we use the following trial function
\begin{align}
w^{(1)}_{ih}(v)=w_{1,1}\cos(v)+w_{1,0},
\end{align}
where $w_{1,1}$ are $w_{1,0}$ constants. By calculating the derivatives from the coefficients we can simply obtain the values of $w_{1,1}$ and $w_{1,0}$, namely
\begin{align}
w_{1,1} = -\dfrac{w_{0,0}\beta}{\alpha+\dfrac{3C^2}{w_{0,0}^4}-1},\qquad w_{1,0}=0.
\end{align}
We use the same steps for the solution of Eq.~\eqref{e:diffeq2}. By using trigonometric identities it can be seen, that the differential equation has the following form
\begin{align}
\hspace{-3mm}
w^{(2)''}(v){+}\left(\alpha{+}\frac{3C^2}{w_{0,0}^4}\right)w^{(2)}(v) {=} \left({-}w_{0,0}\gamma{-}\frac 12w_{1,1}\beta{+}\frac{3C^2w_{1,1}^2}{w_{0,0}^5} \right){-}\left(w_{0,0}\delta{+}\frac 12w_{1,1}\beta{-}\frac{3C^2w_{1,1}^2}{w_{0,0}^5}\right)\cos(2v).
\end{align}
Like in the previous case the solution of the homogeneous part is
\begin{align}
w^{(2)}_h(v) = K_1\sin\left(\sqrt{\alpha+\frac{3C^2}{w_{0,0}^4}}v\right)+K_2\cos\left(\sqrt{\alpha+\frac{3C^2}{w_{0,0}^4}}v\right),
\end{align}
where again $K_1$ and $K_2$ must disappear for the $2\pi$ periodicity, $K_1=K_2\equiv0$. The trial function of the particular solution of the inhomogeneous equation is:
\begin{align}
w^{(2)}_{ih}(v)=w_{2,2}\cos(2v)+w_{2,0}.    
\end{align}
Again by calculating the appropriate derivatives the equality of the coefficients imply:
\begin{align}
w_{2,2} = \dfrac{\dfrac{3C^2w_{1,1}^2}{w_{0,0}^5}-w_{0,0}\delta-\dfrac 12w_{1,1}\beta}{\alpha+\dfrac{3C^2}{w_{0,0}^4}-4},\quad w_{2,0} = \dfrac{\dfrac{3C^2w_{1,1}^2}{w_{0,0}^5}-w_{0,0}\gamma-\dfrac 12 w_{1,1}\beta}{\alpha+\dfrac{3C^2}{w_{0,0}^4}}.
\end{align}
Only the solution of Eq.~\eqref{e:diffeq3} is left
\begin{align}
\begin{split}
&w^{(3)''}(v)+w^{(3)}(v)\left(\alpha+\frac{3C^2}{w_{0,0}^4}\right)=-\Bigg(w_{0,0}\varepsilon+w_{1,1}\gamma +\frac 12 w_{1,1}\delta +w_{2,0}\beta +\frac 12w_{2,2}\beta - \frac{12C^2w_{1,1}w_{2,0}}{w_{0,0}^5}-\\&-\frac{6C^2w_{1,1}w_{2,2}}{w_{0,0}^5}+\frac{15C^2w_{1,1}^3}{2w_{0,0}^6}\Bigg)\cos(v)-\Bigg(w_{0,0}\eta+\frac 12w_{1,1}\delta +\frac 12 w_{2,2}\beta- \frac{6C^2w_{1,1}w_{2,2}}{w_{0,0}^5}+\frac{5C^2w_{1,1}^3}{2w_{0,0}^6} \Bigg)\cos(3v).
\end{split}
\end{align}
The homogeneous solution reads
\begin{align}
w^{(3)}_h(v) = K_1\sin\left(\sqrt{\alpha+\frac{3C^2}{w_{0,0}^4}}v\right)+K_2\cos\left(\sqrt{\alpha+\frac{3C^2}{w_{0,0}^4}}v\right),
\end{align}
where again the constants are $K_1=K_2\equiv0$ due to the periodicity of $w(v)$. The trial function for the particular solution of the inhomogeneous equation
\begin{align}
w^{(3)}_{ih}(v)=w_{3,1}\cos(v)+w_{3,3}\cos(3v),
\end{align}
where the forms for $w_{3,1}$ and $w_{3,3}$ coefficients are
\begin{align}
\begin{split}
w_{3,1}&=-\dfrac{w_{0,0}\varepsilon+w_{1,1}\gamma+\dfrac 12 w_{1,1}\delta +w_{2,0}\beta+\dfrac 12w_{2,2}\beta-\dfrac{12C^2w_{1,1}w_{2,0}}{w_{0,0}^5}-\dfrac{6C^2w_{1,1}w_{2,2}}{w_{0,0}^5}+\dfrac{15C^2w_{1,1}^3}{2w_{0,0}^6}}{\alpha+\dfrac{3C^2}{w_{0,0}^4}-1},\\
w_{3,3}&=-\dfrac{w_{0,0}\eta+\dfrac 12w_{1,1}\delta+\dfrac 12w_{2,2}\beta-\dfrac{6C^2w_{1,1}w_{2,2}}{w_{0,0}^5}+\dfrac{5C^2w_{1,1}^3}{2w_{0,0}^6}}{\alpha+\dfrac{3C^2}{w_{0,0}^4}-9}.
\end{split}
\end{align}
Then by using the fact that $\psi'(v) = Cw^{-2}(v)$, Eq.~(\ref{e:wprime}), $\psi(v)$ can be calculated if we again expand $\psi'(v)$ into Taylor series in $e$ up to third order
\begin{align}
\begin{split}
\hspace{-2mm}\frac 1C\psi(v) &= \frac{v}{w_{0,0}^2}{-}2\frac{w_{1,1}\sin(v)}{w_{0,0}^3}e{+}\frac{1}{w_{0,0}^4}\Bigg\{3w_{1,1}^2\left(\frac{\sin(2v)}{4}{+}\frac v2\right){-}w_{0,0}w_{2,0}v{-}\frac{\sin(2v)w_{0,0}w_{2,2}}{2}\Bigg\}e^2{-}\\
&{-}\frac{1}{w_{0,0}^5}\Bigg\{\frac 23 w_{1,1}^3\left(\frac 94\sin(v){+}\frac{\sin(3v)}{4}\right){+}\sin(v)w_{0,0}^2w_{3,1}{+}\frac{\sin(3v)w_{0,0}^2w_{3,3}}{3}{-}2\sin(v)w_{0,0}w_{1,1}w_{2,0}{-}\\&-2w_{0,0}w_{1,1}w_{2,2}\left(\frac{\sin(v)}{2}+\frac{\sin(3v)}{6}\right)+\frac 23 w_{1,1}\left(\frac{w_{1,1}^2}{4}-\frac{w_{0,0}w_{2,2}}{2}\right)\sin(3v)+\\&+2w_{1,1}\left(\frac 34w_{1,1}^2-\frac 12 w_{0,0}w_{2,2}-w_{0,0}w_{2,0}\right)\sin(v)\Bigg\}e^3+\mathcal O(e^4).
\end{split}
\end{align}
Now we have expressions for $w(v)$ and $\psi(v)$, thus $\xi(v) = aw(v)\cos\big(\psi(v)+b\big)$ can be calculated. One can easily see, that all $w_{i,j}$ coefficients and the expression of $\psi(v)$ have a common multiplicative factor $\sqrt C$ and $C$ respectively, therefore this factor can be chosen to be $C=1$. The effects of $C$ will be considered with the initial conditions. It is left to determine constants $a$ and $b$, which are controlled by the initial conditions $\xi(0)\equiv \xi_0$ and $\xi'(0) \equiv \xi'_0$. As the differential equations are second order linear differential equations with periodic coefficients, the initial conditions can be arbitrary, therefore we use the simple conditions of $x_0 = 1$, $y_0 = 1$, $v_{x0} = 0$, $v_{y0} = 0$, from which $\xi_{1,0}$, $\xi_{1,0}'$, $\xi_{2,0}$ and $\xi_{2,0}'$ can be easily achieved. By using the values $\xi_0$ and $\xi_0'$
\begin{align}
\begin{split}
    \xi_0 &= aw(0)\cos\big(\psi(0)+b\big),\quad \xi_0' =aw'(0)\cos\big(\psi(0)+b\big)-\frac{1}{w(0)}\sin\big(\psi(0)+b\big),\quad\textnormal{therefore}\\
    a &= \sqrt{\big(w'(0)\xi_0-\xi_0'w(0)\big)^2+\left(\frac{\xi_0}{w(0)}\right)^2},\quad b = \arccos\left(\frac{\xi_0}{aw(0)}\right)-\psi(0).
\end{split}
\label{eq:ab_const}
\end{align}
At the end the only task is to use the transformations detailed in Eqs.~\eqref{e:y1}-\eqref{e:y1*}, calculate $y_2$ and $y_2^*$ with Eq.~\eqref{e:y2y2*}, then turn back to the $x,y$ coordinates as $x=y_1+y_1^*$ and $y = y_2+y_2^*$.

\section{Illustrations and discussion}
\label{sec:conclusion}

\begin{figure}[htb]
\centering
\includegraphics[width=0.75\textwidth]{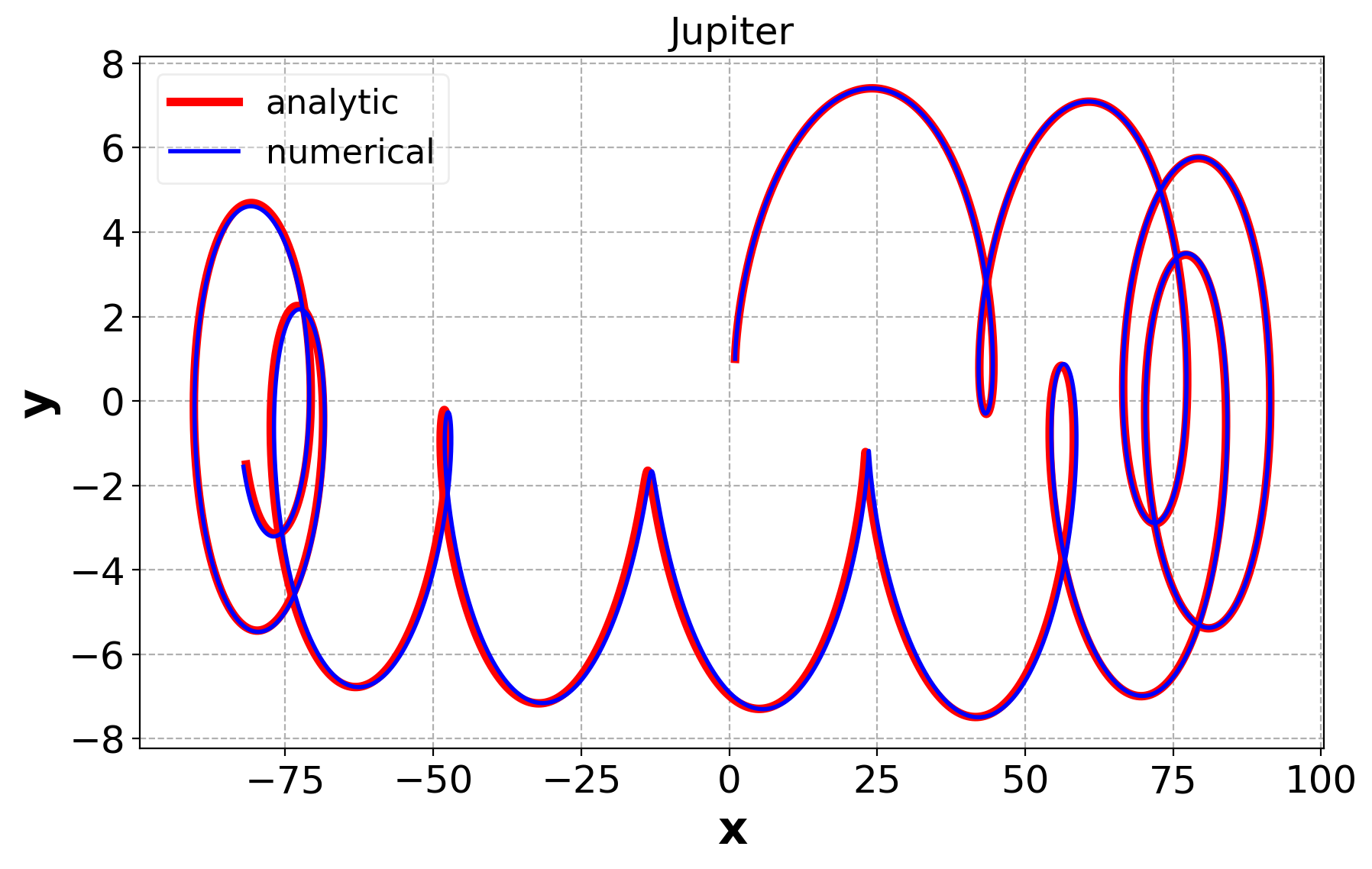}
\caption{Analytic and numerical solution in Sun-Jupiter system. Initial conditions are the same as in Fig~\ref{f:DE_and_Hill_v}.}
\label{f:jupiter}
\end{figure}

The prominent example of co-orbital dynamics is the Sun-Jupiter-Trojan configuration in our own Solar System. We apply the perturbative solution described in Sec.~\ref{sec:series} to this structure first. Fig.~\ref{f:jupiter} depicts the trajectory around the Sun-Jupiter triangular Lagrangian point. The integration time is 20 periods of Jupiter (ca. 240 years). The analytic and numerical solutions match perfectly, although after some time ($\sim 38-40$ periods) they start to deviate.   

\begin{figure}[htb]
\centering
\includegraphics[width=0.495\textwidth]{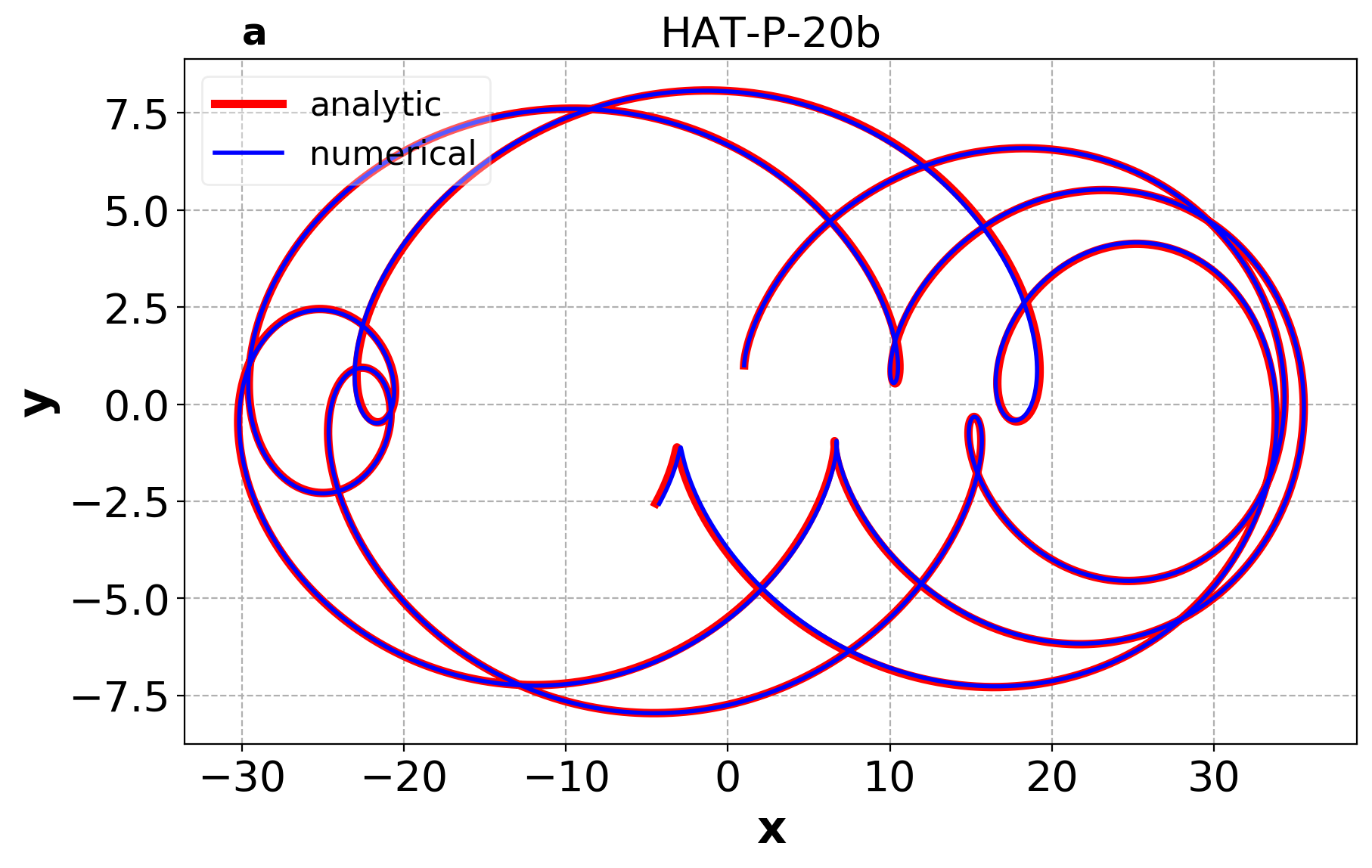}
\includegraphics[width=0.495\textwidth]{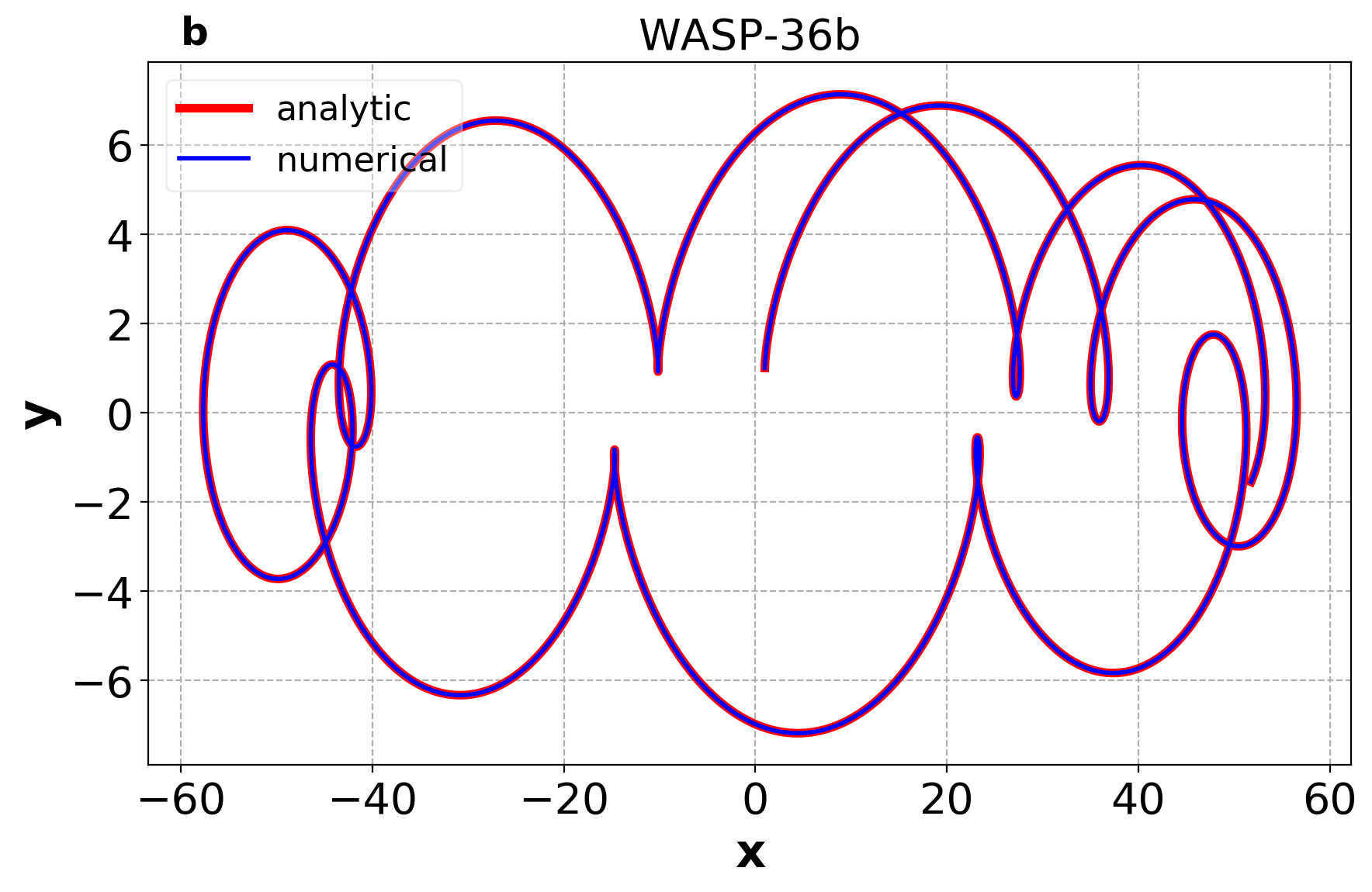}
\caption{Perturbative solution for particular exoplanetary systems. Initial conditions are the same as in Fig~\ref{f:DE_and_Hill_v}. For parameters see the main text.}
\label{f:exoplanets}
\end{figure}

Recently, \cite{LilloBox2018} studied the physical parameters and dynamical properties of possible exo-Trojans in systems with close-in (orbital period $<$ 5 days) planets. We selected two of them, HAT-P-20b ($e=0.015,\;\mu=0.0091$) and WASP-36b ($e=0.0,\;\mu=0.0021$), to demonstrate analytic solution in these regimes\footnote{The orbital periods are HAT-P-20b : 2.87 days, WASP-36b : 1.53 days.}. The orbits are plotted in Fig.~\ref{f:exoplanets}a and b, respectively. The panels show the paths for T=20 periods again. Due to the zero eccentricity of the planet, the analytic solution for WASP-36b remains very close to the numerical outcome for much longer times (not shown here).   

Considering the Earth-Moon system with $e=0.054$ and $\mu=0.012$ it falls close to the limit of third order solution. The analytic solution diverges after 5-6 revolutions ($\sim 130-150$ days) of the Moon. We have seen that for Sun-Jupiter system the analytic curve traces the numerical method reasonable well while the eccentricity falls into the same range. In addition, we have found that the rather large mass parameter - compared to planetary systems - does not affect the precision of the analytic solution provided the eccentricity is small enough, practically zero. This is, however, not the case for Moon. Consequently, systems with moderate non-zero eccentricity and mass parameter in the same magnitude requires further improvement to the analytic solution, e.g. higher order expansion in mass. 

In this work we fully describe the motion around triangular Lagrangian points with Hill's equations. As a perturbative solution, a third order expansion of Floquet function $w(v)$ in eccentricity is presented. This method is capable to follow analytically the orbit of a massless particle around the equilibrium points $\mathrm{L}_4$ and $\mathrm{L}_5$ in ERTBP. Precise trajectory forecast for moderate eccentricity ($e\leq 0.05$) and mass parameter ($\mu \leq 0.005$) is achievable for tens of secondary's orbital period. Furthermore, we note that Eq.~\ref{eq:ab_const} can be used to identify periodic orbits around L$_4$ and L$_5$ points. This calculation is postponed elsewhere.


\begin{acknowledgements}
This work was supported by the NKFIH Hungarian Grants K119993, FK134203. The support of Bolyai Research Fellowship and  \'UNKP-19-2 (BB) and \'UNKP-19-4 (TK) New National Excellence Program of Ministry for Innovation and Technology is also acknowledged.
\end{acknowledgements}

\bibliographystyle{spbasic}      
\bibliography{references}   
\end{document}